      \documentclass[a4paper,11pt]{article}  
\pdfoutput=1 % if your are submitting a pdflatex (i.e. if you have
\usepackage{jheppub1}% for details on the use of the package, please
\usepackage{hyperref} 
\usepackage{cleveref}

\title{\boldmath Asymptotic magnetically charged non-singular black hole and its
thermodynamics}

\author{Askar Ali and Khalid Saifullah} 
\affiliation{Department of Mathematics, Quaid-i-Azam University, Islamabad, Pakistan}
\emailAdd{askarali@math.qau.edu.pk} \emailAdd{ksaifullah@fas.harvard.edu}
\abstract{Some very interesting solutions of the field equations of Einstein's general theory of relativity have been constructed in the framework of nonlinear electrodynamics. In particular, magnetically charged black hole solutions in the framework of exponential nonlinear electrodynamics have been obtained in general relativity in Kruglov (2017) \cite{13}. Using this approach a magnetically charged non-singular black hole spacetime in the framework of
exponential electrodynamics in some modified theory of gravity has been constructed in this letter. The metric describing asymptotic non-singular magnetized black hole is worked out in terms of the parameter
of our model. When this parameter vanishes our solution reduces to the above mentioned black hole solution. Thermodynamics of the resulting solution is also discussed by calculating the Hawking temperature and heat capacity when the magnetic charge is constant. We also find out the point where the first order phase transition induced by temperature changes takes place. The quantum radiations from this black hole are also discussed and the mathematical expression for the rate of energy flux of these radiations has been obtained. 
\vspace{60 mm}
}

\bigskip\bigskip

\notoc

\begin{document}
\maketitle
\flushbottom

%Key Words: Non-linear electrodynamics; Non-singular black holes; Thermodynamics  
%approximation
\section{Introduction}

From the beginning of the last century several proposals 
of nonlinear electrodynamics have been put forth for doing the task to alleviate the singularity
of Maxwell's solution to the field of a point charge. Among these proposals
the formulation of Born and Infeld \cite{1} was very successful.
Non-linear electrodynamics arises when classical electrodynamics is modified
because of quantum corrections. Thus, by assuming loop corrections in QED we get
Heisenberg-Euler electrodynamics \cite{2}, which gives way to the vacuum
birefringence phenomenon. The Born-Infeld electrodynamics does not contain this
phenomenon but in the modified Born-Infeld electrodynamics, this phenomenon comes into
play. Another requirement for any model of nonlinear electrodynamics to be a
successful is that Maxwell's electrodynamics is obtained from it, in the
region where electromagnetic field is weak. There are several desirable
characteristics of Born-Infeld electrodynamics, for instance, electric field has an
upper bound at the centre of charged particles and a finite energy at that
position while in Maxwell's electrodynamics the issue of infinite
electromagnetic energy arises and also the electric field has singularity at the
origin of charged particles. For finiteness of electric field Born and
Infeld chose a particular nonlinear action having maximum field strength;
the field equations thus obtained were coupled with the gravitational field
equations and then they solved the equations to determine a solution corresponding to a point charge which was static and spherically
symmetric. The solutions of the
Einstein field equations coupled to the Born-Infeld electrodynamics were
found \cite{3} in 1937. Other models of nonlinear
electrodynamics \cite{4} have also been used to discuss
the solution of Einstein's equations. Further, Born-Infeld nonlinear
electrodynamics has been coupled with gravity to obtain a class of regular static
spherically symmetric solutions that behave like Reissner-Nordstr\"{o}m
solution asymptotically with respect to a point charge source \cite{5}.  It is to be noted that
classical electrodynamics is modified in case of strong electromagnetic
fields due to the self-interaction of photons \cite{6}.

Another development in this direction came when the solutions describing
black holes in general relativity were studied in the framework of nonlinear
electrodynamics \cite{7, 8, 9, 10, 11, 12}. These solutions asymptotically approach the
Reissner-Nordstr\"{o}m solution at radial infinity. The model of exponential
nonlinear electrodynamics was used to construct magnetically charged
spherically symmetric black hole solution similar to Reissner-Nordstr\"{o}m
solution with a few corrections \cite{13}.

It is well-known that, in both classical and quantum realms, Einstein's theory of gravity is
ultraviolet-incomplete (UV-incomplete). The existence of singularities is
the main problem in this theory, e.g. solutions of Einstein's equations such
as Schwarzschild, Reissner-Nordstr\"{o}m and Kerr metric, have curvature
singularities at the origin. So, in general, one can believe that the
modification of this theory is possible in those regions where the curvature
is very high. Many proposals have been put forward to achieve such modifications. For example, it was also proposed that if higher order terms are included in 
curvature and those terms which contain higher order derivatives, then the theory of gravity can be made UV-complete \cite{14}. However these theories
contain non-physical degrees of freedom, the so-called ghosts. Recently a new UV-complete
modification of the theory of gravity has been put forward in which this problem
does not occur. This theory is called a ghost free gravity theory \cite
{15, 16, 17, 18, 19}. This ghost free theory is also applicable in the problem of
singularities in black holes and cosmology \cite{20, 21, 22, 23, 24}. If the unfailing
fundamental theory is not known then the more naive, phenomenological
approach for description of physics in the regime of high curvature can also
be useful. In this approach one can consider that the gravity is still
described by a classical metric in this regime where the curvature is high.
In view of describing gravity there exists a 
parameter $\mu $ for fundamental energy scale which is related to the fundamental scale length by $l=\mu
^{-1}$. The classical Einstein's equations will be modified if the curvature
is comparable with $l^{-2}$. Instead of correcting the field equations,
we would impose a number of restrictions on the line element. More precisely
we consider: (i) the field equations in the modified theory are
approximately similar to Einstein's equations in the domain where the
curvature is small i.e. $R\ll l^{-2}$; (ii) the metric functions are
regular; (iii) the curvature invariants satisfying the limiting curvature
condition, which means that their value is uniformly constrained by some
fundamental value, $\left\vert R\right\vert \leq kl^{-2}$ \cite{25,26}. Here 
$R$ denotes any type of invariants which can be constructed from curvature
tensor and its covariant derivatives and $k$ is the dimensionless constant.
Those black hole metrics which satisfy the above conditions are called
non-singular black holes. A large number of non-singular black hole models were proposed in Ref. \cite{27}. 

So far, nonlinear electrodynamics was used to determine the asymptotic
charged black hole solutions of Einstein's field equations. Here, in this
work we use the exponential electrodynamics model to find out the asymptotic
magnetically charged non-singular black hole solution of some modified
theory of gravity. We determine the metric and discuss thermodynamics for
this object as well. In order to achieve these goals, we assume a
spherically symmetric metric describing a black hole which is formed when a null shell of mass $M$ collapses. It  exists for a finite
life time and then ends due to the collapse of another shell having mass $-M$. Such a black hole is called a sandwich black hole and is described by the
Hayward metric \cite{28}. Following the approach of Ref. \cite{13} we use the model of exponential electrodynamics on
this metric to work out magnetically charged non-singular black hole
solution. The curvature of the black hole solution discussed in our letter is finite everywhere and is proportional to $l^{-2}$. When we put $l$ equal to zero, our metric reduces to Einstein's theory \cite{13}. When the radial coordinate, $r=0$, these metrics are finite, however, the Kretschmann scalar in our work indicates that it is regular at $r=0$ whereas in Ref. \cite{13} this scalar is infinite which shows the occurrence of true curvature singularity, although the Ricci scalar is non-singular. This shows that the coupling to nonlocal elm provides some but not complete regularization effect. Because of this reason our solution represents a non-singular black hole. Secondly, our solution is asymptotic to charged non-singular black hole in the same manner as the solution in Ref. \cite{13} is asymptotic to Reissner-Nordstr\"{o}m black hole. In this letter the Hawking temperature and heat capacity at constant magnetic charge are also worked out. Apart from Hawking radiations, the non-singular black holes emit quantum radiations also from the interior spacetime, which we have investigated. 

The letter is planned in the following manner. In Section 2 we use the gauge
covariant Dirac quantization of our model coupled with gravity to
derive the metric functions describing a magnetically charged non-singular
black hole solution. The corrections to this black hole are found at radial
infinity. We also calculate the curvature scalar invariants and their
asymptotic behaviour at $r\rightarrow \infty $ and at $r\rightarrow 0$. Section 3 deals with thermodynamics of the magnetically charged non-singular black hole
is studied. It is also shown here that not only the first order but the second order phase
transitions also take place in such objects. In Section 4 we study quantum radiations from our resulting solution. In this section we also find
formulae for the gain function and for quantum energy flux in the case of magnetically
charged non-singular black hole. We summarize our results in Section 5.  

\section{Asymptotic magnetically charged non-singular black hole solution}

The solutions describing black holes in general relativity within the
framework of nonlinear electrodynamics have been studied \cite{7, 8, 9, 10, 11, 12}. These
solutions describe electrically charged black holes which asymptotically
approach Reissner-Nordstr\"{o}m solution at radial infinity. In order to
described magnetically charged black hole solutions we consider the
exponential nonlinear electrodynamics \cite{13} for which the Lagrangian density is given by 
\begin{equation}
\pounds =-P\exp (-\beta P),  \label{A1}
\end{equation}%
where $P=\left( 1/4\right) F_{\mu \nu }F^{\mu \nu }=\left( \mathbf{B}^{2}-%
\mathbf{E}^{2}\right) /2,~F^{\mu \nu }=\partial ^{\mu }A^{\nu }-\partial
^{\nu }A^{\mu }.$ Here $F^{\mu \nu }$ is the electromagnetic field tensor, $%
A^{\mu }$ is the four-potential, $\mathbf{B}$ is the magnetic field, $%
\mathbf{E}$ is the electric field, and $\beta $ is the parameter which has
dimensions of (length)$^{4}$ and its upper bound $\left( \beta \leq 1\times
10^{-23}\text{T}^{-2}\right) $ which was found from PVLAS experiment. Now, the Euler-Lagrange equations are%
\begin{equation*}
\partial _{\mu }\left( \frac{\partial \pounds }{\partial \left( \partial
_{\mu }A_{v}\right) }\right) -\frac{\partial \pounds }{\partial A_{v}}=0,%
\text{ \ where }\mu ,v=0,1,2,3.
\end{equation*}%
Thus the field equations become 
\begin{equation}
\partial _{\mu }\left[ \left( \beta P-1\right) \exp \left( -\beta P\right)
F^{\mu \nu }\right] =0.  \label{A2}
\end{equation}%
The energy-momentum tensor is given by 
\begin{equation}
\tau ^{\mu \upsilon }=H^{\mu \lambda }F_{\lambda }^{\upsilon }-g^{^{\mu
\upsilon }}\pounds ,  \label{A3}
\end{equation}%
where $g^{^{\mu \upsilon }}$ is the reciprocal metric tensor and the
quantity $H^{\mu \lambda }$ is given by 
\begin{equation}
H^{\mu \lambda }=\frac{\partial \pounds }{\partial F_{\mu \lambda }}=-\left(
1-\beta P\right) \exp \left( -\beta P\right) F^{\mu \lambda }.  \label{A4}
\end{equation}%
Thus we can find the energy-momentum tensor from the Lagrangian density (\ref%
{A1}) as%
\begin{equation}
\tau ^{\mu \upsilon }=\exp \left( -\beta P\right) \left[ \left( \beta
P-1\right) F^{\mu \lambda }F_{\lambda }^{\upsilon }+g^{\mu \upsilon }P\right]
,  \label{A5}
\end{equation}%
from which its trace can be calculated as 
\begin{equation}
\tau =4\beta P^{2}\exp \left( -\beta P\right) .  \label{A6}
\end{equation} 
For weak fields, or when $\beta \rightarrow 0$, we get the results of
classical electrodynamics, i.e., $\pounds \rightarrow -P$ and the trace of
the energy-momentum tensor becomes zero. In general, $\beta \neq 0$, and the
non-zero trace of the energy-momentum tensor means that the scale invariance
is violated in the theory. So, any variants of nonlinear electrodynamics
with the dimensional parameter give the breaking of scale invariance and so
the divergence of dilation current does not vanish, i.e., $\partial
_{v}D^{\upsilon }=\tau $ where $D^{\upsilon }=x^{\mu }\tau _{\mu }^{\upsilon}$.

If the general principles of causality and unitarity hold here then the
theory is workable. According to this principle the group velocity of
excitations over the background does not exceed the speed of light. This
gives the requirement \cite{11, 12, 13} that $\pounds _{P}=\partial \pounds /\partial P 
\leq 0$. So, from Eq. (\ref{A1}) we get to the point that
when $\beta P\leq 1$ the causality principle holds. In the case of pure
magnetic field we have the condition $B\leq \sqrt{\frac{2}{\beta }}.$ The
unitarity principle holds when $\pounds _{P}+2P\pounds _{PP}\leq 0$ and $%
\pounds _{PP}\geq 0$, and thus with the help of Eq. (\ref{A1}) we get $\beta
P\leq 0.219$ which is the restriction for unitarity principle \cite{13}. So,
causality and unitarity both take place when $\beta P\leq 0.219$ and for
purely magnetic field this yields the requirement 
\begin{equation*}
\mathbf{B}\leq \sqrt{\frac{5-\sqrt{17}}{2\beta }}\simeq \frac{0.66}{\sqrt{%
\beta }}.
\end{equation*}

Now, we derive the metric which represents the static
magnetic non-singular black hole. The invariant $P$ for pure magnetic field
in the spherically symmetric spacetime, is given as 
\begin{equation}
P=\frac{q^{2}}{2r^{4}}.  \label{A7}
\end{equation}%
The most general spherically symmetric spacetime is defined by 

\begin{equation*} 
dS^{2} =\sigma ^{2}ds^{2},    %\notag   
\end{equation*}%
where 
\begin{equation}
ds^{2} =-\alpha ^{2}fdv^{2}+2\alpha dvdr+r^{2}\left( d\theta ^{2}+\sin
^{2}\theta d\phi ^{2}\right) .   \label{A8} 
\end{equation} 
Here $\sigma $ is the conformal factor and $\alpha $ and $f$ \ are functions
of radial coordinate $r$. If we put 
\begin{equation}
\alpha =1,\text{ \ \ }f=1-\frac{2Mr^{2}}{r^{3}+2Ml^{2}},  \label{A9}
\end{equation}%
then Eq. (\ref{A8}) represents the Hayward metric \cite{28} which describes
uncharged static non-singular black hole and is the solution of some
modified theory of gravity. The non-singular black hole defined by
this line element is also called a sandwich black hole. In the assumption that
the mass $M$ of this black hole varies with $r$, we can write 

\begin{equation}
M(r)=\int_{0}^{r}\rho (r)r^{2}dr=m-\int_{r}^{\infty }\rho
(r)r^{2}dr.  \label{A10}
\end{equation} 
In the above equation $m=\int_{0}^{\infty }\rho (r)r^{2}dr$
represents the black hole's magnetic mass. The energy density, in the case
of zero electric field, can be written from Eq. (\ref{A5}) 
\begin{equation}
\rho =\frac{q^{2}}{2r^{4}}\exp \left( \frac{-\beta q^{2}}{2r^{4}}\right) .
\label{A11}
\end{equation}%
Thus the mass function becomes%
\begin{equation}
M(r)=\frac{q^{2}}{2}\int_{0}^{r}\exp \left( \frac{-\beta q^{2}}{%
2r^{4}}\right) \frac{dr}{r^{2}}.  \label{A12}
\end{equation}%
Or, using the incomplete gamma function%
\begin{equation}
\Gamma (s,x)=\int_{x}^{\infty }t^{s-1}e^{-t}dt,  \label{A13}
\end{equation}%
this takes the form 
\begin{subequations}
\begin{equation}
M(r)=\frac{q^{\frac{3}{2}}\Gamma \left( \frac{1}{4},\frac{\beta q^{2}}{2r^{4}%
}\right) }{2^{\frac{11}{4}}\beta ^{\frac{1}{4}}}.  \label{A14}
\end{equation}%
The magnetic mass of the black hole is then given by 
\end{subequations}
\begin{equation}
m=M(\infty )=\frac{q^{\frac{3}{2}}\Gamma \left( \frac{1}{4}\right) }{2^{%
\frac{11}{4}}\beta ^{\frac{1}{4}}}\simeq \frac{0.54q^{\frac{3}{2}}}{\beta ^{%
\frac{1}{4}}}.  \label{A15}
\end{equation}%
Thus the metric function becomes%
\begin{equation}
f\left( r\right) =1-\frac{r^{2}q^{\frac{3}{2}}\Gamma \left( \frac{1}{4},%
\frac{\beta q^{2}}{2r^{4}}\right) 2^{\frac{-7}{4}}\beta ^{\frac{-1}{4}}}{%
r^{3}+l^{2}q^{\frac{3}{2}}2^{\frac{-7}{4}}\beta ^{\frac{-1}{4}}\Gamma \left( 
\frac{1}{4},\frac{\beta q^{2}}{2r^{4}}\right) }.  \label{H1}
\end{equation}%
If we put $l=0$, we obtain the metric function for Einstein's theory \cite{13} . Using the above results we can write the asymptotic value of the
metric function in the neighbourhood of radial infinity. For this we use the series expansion 
\begin{subequations}
\begin{equation}
\Gamma (s,z)=\Gamma (s)-z^{s}\left[ \frac{1}{s}-\frac{z}{s+1}+\frac{z^{2}}{%
2\left( s+2\right) }+O\left( z^{3}\right) \right] ,\text{ \ }z\rightarrow 0.
\label{A16}
\end{equation}%
Thus the metric function $f\left( r\right) $ at $r\rightarrow \infty $ takes
the following form 
\end{subequations}
\begin{equation}
f\left( r\right) =1-\frac{r^{2}\left[ 2m-\frac{q^{2}}{r}+\frac{\beta q^{4}}{%
20r^{5}}+O(r^{-9})\right] }{r^{3}+l^{2}\left[ 2m-\frac{q^{2}}{r}+\frac{\beta
q^{4}}{20r^{5}}+O(r^{-9})\right] }.  \label{A17}
\end{equation}%
In the numerator and denominator if we choose $\beta =0$ i.e. by neglecting
the nonlinear effects of magnetic field we get%
\begin{equation}
f\left( r\right) =1-\frac{\left( 2mr-q^{2}\right) r^{2}}{r^{4}+l^{2}\left(
2mr-q^{2}\right) },  \label{A18}
\end{equation}%
which corresponds to the metric of charged non-singular black hole \cite{28}%
. In the limit $l\rightarrow 0,$ a metric similar to the Reissner-Nordstr%
\"{o}m solution is obtained. Further, from (\ref{A17}) we see that%
\begin{equation}
f\left( r\right) \approx 1-\frac{2m}{r}+\frac{q^{2}}{r^{2}}+\frac{4m^{2}l^{2}%
}{r^{4}}-\frac{\left( 40q^{2}l^{2}+\beta q^{4}\right) }{r^{5}}.  \label{A19}
\end{equation}%
By making corrections to the fourth order terms, a metric similar to the
Reissner-Nordstr\"{o}m solution is obtained. The limit $r\rightarrow \infty$, gives Minkowski spacetime.

We can also write the asymptotic values of the metric function at $r\rightarrow
0 $. For doing this we will use the series expansion 
\begin{equation}
\Gamma (s,z)=\exp (-z)z^{s}\left[ \frac{1}{z}+\frac{s-1}{z^{2}}+\frac{%
s^{2}-3s+2}{z^{3}}+O(z^{-4})\right] ,\text{ \ }z\rightarrow \infty ,
\label{A20}
\end{equation}%
and obtain the expression 
\begin{equation}
f\left( r\right) \approx 1+\exp \left( \frac{-\beta q^{2}}{2r^{4}}\right) %
\left[ \frac{-l^{2}}{2\beta }-\frac{r^{2}}{2\beta }+\frac{3l^{2}r^{4}}{%
4\beta ^{2}q^{2}}+\frac{3r^{6}}{4\beta ^{2}q^{2}}\right] .  \label{A21}
\end{equation}%
The above result shows that the metric function is finite at the origin. Let
us define here a new variable $x$ in terms of the radial coordinate $r$, by 
\begin{equation}
x=\left( \frac{2}{\beta q^{2}}\right) ^{\frac{1}{4}}r,  \label{A22}
\end{equation}%
so that (\ref{A18}) becomes%
\begin{equation}
f(x)=1-\frac{\Gamma (\frac{1}{4},\frac{1}{x^{4}})qx^{2}\beta ^{\frac{1}{2}}}{%
2^{\frac{3}{2}}x^{3}\beta +\sqrt{2}l^{2}\Gamma (\frac{1}{4},\frac{1}{x^{4}})}%
.  \label{A23}
\end{equation} 

\begin{figure}[!h]
  \centering
  \includegraphics[width=0.7\textwidth]{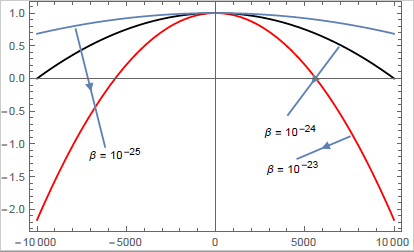}
  \caption{Three plots of function $f(x)$ for fixed values of $Q=1$ and $l=0.01$, with different values of $\protect\beta$ for each curve.}
  \label{FigOne}
  \end{figure}

\begin{figure}[!h]
  \centering
  \includegraphics[width=0.7\textwidth]{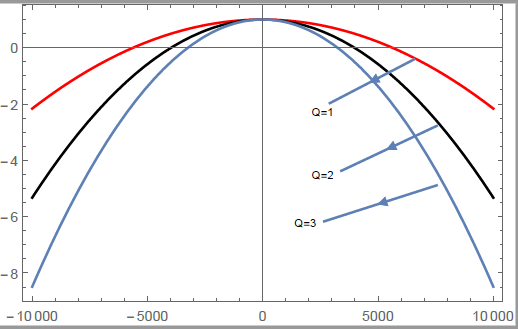}
  \caption{Three plots of function $f(x)$ for fixed values of $\protect\beta =1\times 10^{-23}$ and $l=0.01$, with
different values of $Q$ for each curve.}
  \label{FigTwo}
  \end{figure} 
  
\begin{figure}[!h]
  \centering
  \includegraphics[width=0.7\textwidth]{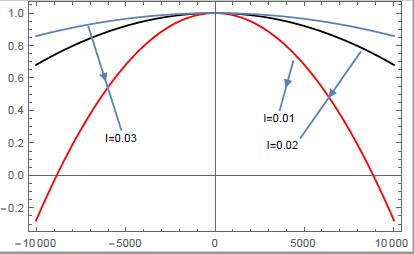}
  \caption{Three plots of function $f(x)$ for fixed values of $\protect\beta =1\times 10^{-25}$ and $Q=0.01$, with
different values of $l$ for each curve.}
  \label{FigThree}
  \end{figure}

Clearly it can be seen that apparent horizons can be found by solving
equation $f(x)=0$. The function given by Eq. (\ref{A23}) is plotted in Figs. 1-3 for different values of $\beta $, $l$ and $Q=q/\sqrt{2}$. The points where these curves intersect horizontal axes indicate the position of the
apparent horizon.

Now, we want to confirm that our metric is asymptotically flat and regular at 
$r=0$, i.e., we have indeed a non-singular object under discussion. For this
purpose the Ricci scalar is given by \cite{27}%
\begin{equation}
R=\frac{d^{2}f}{dr^{2}}+\frac{4}{r}\frac{df}{dr}-2\frac{f-1}{r^{2}}.
\label{A24}
\end{equation}%
From the quadratic invariant $C^{2}=C_{\mu \upsilon \alpha \beta }C^{\mu
\upsilon \alpha \beta }$, where $C_{\mu \upsilon \alpha \beta }$ is the Weyl
tensor, we obtain 
\begin{equation}
C=\frac{1}{\sqrt{3}}\left[ \frac{d^{2}f}{dr^{2}}-\frac{2}{r}\frac{df}{dr}+2%
\frac{f-1}{r^{2}}\right] .  \label{A24a}
\end{equation}%
Differentiating (\ref{A23}) we obtain%
\begin{equation}
\frac{df}{dx}=\frac{-q\left[ 8x^{3}\beta ^{\frac{3}{2}}\exp \left( -\frac{1}{%
x^{4}}\right) +2x\beta ^{\frac{1}{2}}l^{2}\Gamma \left( \frac{1}{4},\frac{1}{%
x^{4}}\right) ^{2}-2x^{4}\beta ^{\frac{3}{2}}\Gamma \left( \frac{1}{4},\frac{%
1}{x^{4}}\right) \right] }{2^{\frac{5}{2}}x^{6}\beta ^{2}+\sqrt{2}%
l^{4}\Gamma \left( \frac{1}{4},\frac{1}{x^{4}}\right) ^{2}+2^{\frac{5}{2}%
}l^{2}\beta x^{3}\Gamma \left( \frac{1}{4},\frac{1}{x^{4}}\right) }.
\label{A25}
\end{equation}%
Differentiating again gives%
\begin{equation}
\frac{d^{2}f}{dx^{2}}=-q\sqrt{\frac{\beta }{2}}\frac{H\left( x\right)
+L\left( x\right) +N\left( x\right) }{S\left( x\right) +W\left( x\right) },
\label{A26}
\end{equation}%
where $H,~L,~N,~S~$and $W~$are given by%
\begin{equation}
H\left( x\right) =16\beta ^{3}x^{9}\Gamma \left( \frac{1}{4},\frac{1}{x^{4}}%
\right) -128\beta ^{3}x^{8}\exp \left( -\frac{1}{x^{4}}\right) -48l^{2}\beta
^{2}x^{6}\Gamma \left( \frac{1}{4},\frac{1}{x^{4}}\right) ^{2},  \label{A27}
\end{equation}%
\begin{equation}
L\left( x\right) =2l^{6}\Gamma \left( \frac{1}{4},\frac{1}{x^{4}}\right)
^{4}-64l^{4}\beta x\exp \left( -\frac{2}{x^{4}}\right) \Gamma \left( \frac{1%
}{4},\frac{1}{x^{4}}\right) +64l^{4}\beta x^{2}\exp \left( -\frac{1}{x^{4}}%
\right) \Gamma \left( \frac{1}{4},\frac{1}{x^{4}}\right)^{2},  \label{A28}
\end{equation}%
\begin{equation}
N\left( x\right) =64l^{2}\beta ^{2}x^{5}\exp \left( -\frac{1}{x^{4}}\right)
\Gamma \left( \frac{1}{4},\frac{1}{x^{4}}\right) -128l^{2}\beta
^{2}x^{4}\exp \left( -\frac{2}{x^{4}}\right) -24l^{4}\beta x^{3}\Gamma
\left( \frac{1}{4},\frac{1}{x^{4}}\right)^{3},  \label{A29}
\end{equation}%
\begin{equation}
S\left( x\right) =16\beta ^{4}x^{12}+l^{8}\Gamma \left( \frac{1}{4},\frac{1}{%
x^{4}}\right) ^{4}+24l^{4}\beta ^{2}x^{6}\Gamma \left( \frac{1}{4},\frac{1}{%
x^{4}}\right)^{2},  \label{A30}
\end{equation}%
\begin{equation}
W\left( x\right) =32l^{2}\beta ^{3}x^{9}\Gamma \left( \frac{1}{4},\frac{1}{%
x^{4}}\right) +8l^{6}\beta x^{3}\Gamma \left( \frac{1}{4},\frac{1}{x^{4}}%
\right)^{3}.  \label{A31}
\end{equation}%
Using the expansion%
\begin{equation}
\Gamma \left( \frac{1}{4},\frac{1}{x^{4}}\right) =\Gamma \left( \frac{1}{4}%
\right) -\frac{1}{x}\left[ 4-\frac{4}{5x^{5}}+O\left( x^{-8}\right) \right]
,~~~x\rightarrow \infty ,  \label{A32}
\end{equation}%
in the expression of Ricci scalar we note that 
\begin{equation}
\lim_{r\rightarrow \infty }R(r)=0.  \label{A33}
\end{equation}%
Similarly 
\begin{equation}
\lim_{r\rightarrow \infty }C(r)=0.  \label{A34}
\end{equation} 
By using the series expansion 
\begin{equation}
\Gamma \left( \frac{1}{4},\frac{1}{x^{4}}\right) =\exp \left( -\frac{1}{x^{4} 
}\right) \left[ x^{3}-\frac{3}{4}x^{7}+O\left( x^{11}\right) \right]
,~~x\rightarrow 0,  \label{A35}
\end{equation} 
we conclude that 
\begin{equation}
\lim_{r\rightarrow 0}R(r)=0,  \label{A36}
\end{equation} 
and 
\begin{equation}
\lim_{r\rightarrow 0}C(r)=0.  \label{A37}
\end{equation} 
This clearly shows that the scalar curvature has no singularities. The
spacetime becomes Minkowski at $r\rightarrow \infty$, while at the origin $r=0$, finite curvature suggests that the black hole under consideration is regular. The Kretschmann scalar is regular at $r=0$, which indicates that our solution of some modified theory of gravity describes a non-singular black hole with exponential pure magnetic source. This is in contrast to Einstein's theory in the framework of exponential electrodynamics \cite{13} where the Kretschmann scalar is singular at $r=0$. It is worth mentioning that as the charged generalization of the Hayward metric in Maxwell's electrodynamics is non-singular \cite{28}, our solution in nonlinear electrodynamics also describes a non-singular black hole, where the curvature of the spacetime is finite everywhere.

\section{Thermodynamics of magnetically charged non-singular black hole}

Here we want to investigate the thermal stability of magnetized non-singular black
hole by working out the Hawking temperature and its heat capacity. The black
hole is unstable where the temperature becomes negative. The Hawking
temperature is described by the relation \cite{SH, GS, RS} 
\begin{equation}
T_{H}=\frac{\kappa }{2\pi },  \label{c1}
\end{equation}%
where $\kappa $ defines the surface gravity which is given by 
\begin{equation}
\kappa =\left. \frac{1}{2}\frac{df}{dr}\right\vert _{H}.  \label{c2}
\end{equation}%
Thus if $r_{1}$ and $r_{2}$ are the inner and outer apparent horizons,
respectively, then for the inner horizon the surface gravity is%
\begin{equation}
\kappa _{1}=\left. \frac{1}{2}\frac{df}{dr}\right\vert _{r_{1}}=\frac{1}{2}%
\left. \frac{df}{dx}\frac{dx}{dr}\right\vert _{x_{1}}.  \label{c3}
\end{equation}%
By using the relation (\ref{A25}) in the above we get%
\begin{equation}
\kappa _{1}=\frac{\sqrt{q}\beta ^{\frac{1}{4}}}{2^{\frac{5}{4}}}\frac{\left[
2x_{1}^{4}\beta \Gamma \left( \frac{1}{4},\frac{1}{x_{1}^{4}}\right)
-8x_{1}^{3}\beta \exp \left( \frac{-1}{x_{1}^{4}}\right) -2x_{1}l^{2}\Gamma
\left( \frac{1}{4},\frac{1}{x_{1}^{4}}\right) ^{2}\right] }{4x_{1}^{6}\beta
^{2}+l^{4}\Gamma \left( \frac{1}{4},\frac{1}{x_{1}^{4}}\right)
^{2}+4l^{2}\beta x_{1}^{3}\Gamma \left( \frac{1}{4},\frac{1}{x_{1}^{4}}%
\right) }.  \label{c4}
\end{equation}%
Similarly, for the outer horizon the surface gravity is given by%
\begin{equation}
\kappa _{2}=\frac{\sqrt{q}\beta ^{\frac{1}{4}}}{2^{\frac{5}{4}}}\frac{\left[
2x_{2}^{4}\beta \Gamma \left( \frac{1}{4},\frac{1}{x_{2}^{4}}\right)
-8x_{2}^{3}\beta \exp \left( \frac{-1}{x_{2}^{4}}\right) -2x_{2}l^{2}\Gamma
\left( \frac{1}{4},\frac{1}{x_{2}^{4}}\right) ^{2}\right] }{4x_{2}^{6}\beta
^{2}+l^{4}\Gamma \left( \frac{1}{4},\frac{1}{x_{2}^{4}}\right)
^{2}+4l^{2}\beta x_{2}^{3}\Gamma \left( \frac{1}{4},\frac{1}{x_{2}^{4}}%
\right) },  \label{c5}
\end{equation}%
so that the expression of Hawking temperature yields the result%
\begin{equation}
T_{H}=\frac{\sqrt{q}\beta ^{\frac{1}{4}}}{2^{\frac{5}{4}}\pi }\frac{\left[
x_{2}^{4}\beta \Gamma \left( \frac{1}{4},\frac{1}{x_{2}^{4}}\right)
-4x_{2}^{3}\beta \exp \left( \frac{-1}{x_{2}^{4}}\right) -x_{2}l^{2}\Gamma
\left( \frac{1}{4},\frac{1}{x_{2}^{4}}\right) ^{2}\right] }{4x_{2}^{6}\beta
^{2}+l^{4}\Gamma \left( \frac{1}{4},\frac{1}{x_{2}^{4}}\right)
^{2}+4l^{2}\beta x_{2}^{3}\Gamma \left( \frac{1}{4},\frac{1}{x_{2}^{4}}%
\right) }.  \label{c6}
\end{equation}

\begin{figure}[!h]
  \centering
  \includegraphics[width=0.7\textwidth]{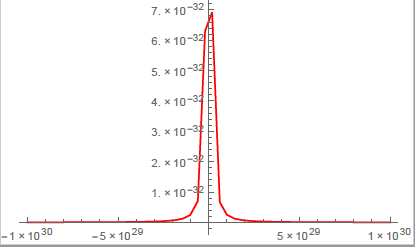}
  \caption{The curve of Hawking temperature for fixed $\protect\beta =1\times 10^{-26}$ and $l=0.02$.}
  \label{FigFour}
  \end{figure}

%\begin{equation*}
%\FRAME{itbpFU}{4.3742in}{2.6143in}{0in}{\Qcb{Fig. 4: The curve of Hawking
%temperature for fixed $\protect\beta =1\times 10^{-26}$ and $l=0.02$. }}{}{%
%askar4.png}{\special{language "Scientific Word";type
%"GRAPHIC";maintain-aspect-ratio TRUE;display "USEDEF";valid_file "F";width
%4.3742in;height 2.6143in;depth 0in;original-width 4.3232in;original-height
%2.5728in;cropleft "0";croptop "1";cropright "1";cropbottom "0";filename
%'Askar4.PNG';file-properties "XNPEU";}}
%\end{equation*} 

This reduces to the result for the black hole solution of Einstein's theory with exponential magnetic source \cite{13}, if we put $l=0$. From Fig. 4 it is clear that the first order phase transition of black hole
occurs at $\pm 5\times 10^{29}$ because the Hawking temperature is zero
there. Hawking temperature gives the maximum value as $x_{2}\rightarrow 0$. The
entropy of the black hole satisfies Hawking area law i.e. $S=A_{h}/4
=\pi r_{2}^{2}$. Then heat capacity is defined for constant charge as%
\begin{equation}
C_{q}=\left. T_{H}\frac{\partial S}{\partial T_{H}}\right\vert _{q}=T_{H}%
\frac{\partial S/\partial r_{2}}{\partial T_{H}/\partial r_{2}}=\frac{2\pi
r_{2}T_{H}}{\partial T_{H}/\partial r_{2}}.  \label{c7}
\end{equation}%
Thus with the help of Eq. (\ref{c6}) we get the expression for heat capacity
in the form%
\begin{equation}
C_{q}=\frac{8\pi x_{2}q\sqrt{\beta }\left[ A\left( x_{2}\right) +B\left(
x_{2}\right) +\beta x_{2}^{3}\exp \left( \frac{-1}{x_{2}^{4}}\right) C\left(
x_{2}\right) \right] }{2^{\frac{3}{2}}D\left( x_{2}\right) \left[ E\left(
x_{2}\right) +64\beta x_{2}\exp \left( \frac{-1}{x_{2}^{4}}\right) F\left(
x_{2}\right) \right] },  \label{c8}
\end{equation}%
where the functions $A$,$~B$,~$C$,~$D$, $E$ and $F$ are given by%
\begin{eqnarray}
A\left( x_{2}\right)  &=&\left. 16l^{2}\beta ^{4}x_{2}^{13}\Gamma \left( 
\frac{1}{4},\frac{1}{x_{2}^{4}}\right) \left[ \Gamma \left( \frac{1}{4},%
\frac{1}{x_{2}^{4}}\right) -2\right] +8l^{6}\beta ^{2}x_{2}^{7}\Gamma \left( 
\frac{1}{4},\frac{1}{x_{2}^{4}}\right) ^{3}\left[ 3\Gamma \left( \frac{1}{4},%
\frac{1}{x_{2}^{4}}\right) -1\right] \right.   \notag \\
&&\left. +l^{8}\beta x_{2}^{4}\Gamma \left( \frac{1}{4},\frac{1}{x_{2}^{4}}%
\right) ^{4}\left[ 8\Gamma \left( \frac{1}{4},\frac{1}{x_{2}^{4}}\right) -1%
\right] ,\right.   \label{c9}
\end{eqnarray}%
\begin{equation}
B\left( x_{2}\right) =l^{10}x_{2}\Gamma \left( \frac{1}{4},\frac{1}{x_{2}^{4}%
}\right) ^{6}-16\beta ^{5}x_{2}^{16}+32l^{4}\beta ^{3}x_{2}^{10}\Gamma
\left( \frac{1}{4},\frac{1}{x_{2}^{4}}\right) ^{2}\left[ \Gamma \left( \frac{%
1}{4},\frac{1}{x_{2}^{4}}\right) -1\right] ,  \label{c10}
\end{equation}%
\begin{eqnarray}
C\left( x_{2}\right)  &=&64\beta ^{4}x_{2}^{12}+4l^{8}\Gamma \left( \frac{1}{%
4},\frac{1}{x_{2}^{4}}\right) ^{4}+96l^{4}\beta ^{2}x_{2}^{6}\Gamma \left( 
\frac{1}{4},\frac{1}{x_{2}^{4}}\right) ^{2}+128l^{2}\beta
^{3}x_{2}^{9}\Gamma \left( \frac{1}{4},\frac{1}{x_{2}^{4}}\right)   \notag \\
&&+32\beta l^{6}x_{2}^{3}\Gamma \left( \frac{1}{4},\frac{1}{x_{2}^{4}}%
\right) ^{3},  \label{c11}
\end{eqnarray}%
\begin{equation}
D\left( x_{2}\right) =4x_{2}^{6}\beta ^{2}+l^{4}\Gamma \left( \frac{1}{4},%
\frac{1}{x_{2}^{4}}\right) ^{2}+4l^{2}\beta x_{2}^{3}\Gamma \left( \frac{1}{4%
},\frac{1}{x_{2}^{4}}\right) ,  \label{c12}
\end{equation}%
\begin{equation}
E\left( x_{2}\right) =2l^{6}\Gamma \left( \frac{1}{4},\frac{1}{x_{2}^{4}}%
\right) ^{4}+16\beta ^{3}x_{2}^{9}\Gamma \left( \frac{1}{4},\frac{1}{%
x_{2}^{4}}\right) -48l^{2}\beta ^{2}x_{2}^{6}-24\beta l^{4}x_{2}^{3}\Gamma
\left( \frac{1}{4},\frac{1}{x_{2}^{4}}\right) ^{3},  \label{c13}
\end{equation} 

\begin{eqnarray}
F\left( x_{2}\right) &=&-2\beta ^{2}x_{2}^{7}+l^{2}\beta x_{2}^{4}\Gamma
\left( \frac{1}{4},\frac{1}{x_{2}^{4}}\right) -2l^{2}\beta x_{2}^{3}\exp
\left( \frac{-1}{x_{2}^{4}}\right) -l^{4}x_{2}\Gamma \left( \frac{1}{4},%
\frac{1}{x_{2}^{4}}\right) ^{2}   \notag \\
&&-l^{4}\exp \left( \frac{-1}{x_{2}^{4}}\right)
\Gamma \left( \frac{1}{4},\frac{1}{x_{2}^{4}}\right) .  \label{c14}
\end{eqnarray}

\begin{figure}[!h]
  \centering
  \includegraphics[width=0.7\textwidth]{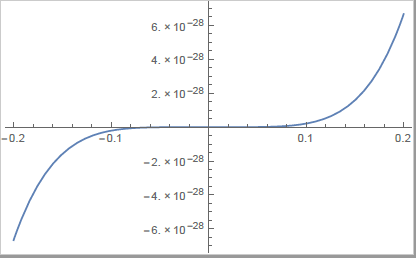}
  \caption{The graph representing heat capacity for fixed values of $\protect\beta $ and $l$. }
  \label{FigFive}
  \end{figure}

%\begin{equation*}
%\FRAME{itbpFU}{4.3846in}{2.7302in}{0in}{\Qcb{Fig. 5: The graph representing
%heat capacity for fixed values of $\protect\beta $ and $l$. }}{}{askar5.png}{%
%\special{language "Scientific Word";type "GRAPHIC";maintain-aspect-ratio
%TRUE;display "USEDEF";valid_file "F";width 4.3846in;height 2.7302in;depth
%0in;original-width 4.3336in;original-height 2.6878in;cropleft "0";croptop
%"1";cropright "1";cropbottom "0";filename 'Askar5.PNG';file-properties
%"XNPEU";}}
%\end{equation*} 

For fixed values of $\beta =1\times 10^{-26}$ and $l=0.02$, the graph in
Fig. 5 of $2^{\frac{3}{2}}C_{q}/8\pi q\sqrt{\beta }$ vs $x_{2}$ shows that
heat capacity is singular in the interval $(-0.05969,0.05969)$, which says
that second order phase transition occurs in this interval. For all values
of $x_{+}<-0.05969$ the heat capacity becomes negative and so the black hole is unstable.
Fig. 5 shows that heat capacity diverges as $T_{H}$ increases.

\section{Quantum radiations from magnetically charged non-singular black hole}

When a black hole is formed, it emits quantum radiation.
An observer outside the black hole can register outgoing Hawking radiation
from the black hole, due to which its mass decreases and
it shrinks in size. One possibility is that the black hole disappears completely
as a result of evaporation. For a spherically symmetric spacetime of
non-singular black hole, this implies that the apparent horizon is closed, and
there is no event horizon. Thus, according to the usual
definition this object, in fact, is not a black hole, but its long-time analogue. However, we are using the same term black hole for these objects too. The
most important property of such type of non-singular black holes is that not
only Hawking radiation is emitted but in addition to that 
quantum radiation also comes from the interior spacetime \cite{28}. We should expect that after the complete evaporation of this object,
the total energy loss by it will be equal to its initial mass. Since, in
this letter we coupled the non-singular black hole solution with the model of
exponential nonlinear electrodynamics, so we consider the model
\begin{equation}
ds^{2}=-fdv^{2}+2dvdr+r^{2}d\omega ^{2},  \label{d1}
\end{equation}%
where the function $f\left( r\right)$ is given by 
\begin{equation}
f\left( r\right) =1-\frac{r^{2}q^{\frac{3}{2}}\Gamma \left( \frac{1}{4},%
\frac{\beta q^{2}}{2r^{4}}\right) }{2^{\frac{7}{4}}\beta ^{\frac{1}{4}%
}r^{3}+l^{2}q^{\frac{3}{2}}\Gamma \left( \frac{1}{4},\frac{\beta q^{2}}{%
2r^{4}}\right) }.  \label{d2}
\end{equation}%
This function depends on $r$ for some real interval $0<v<p$, while the
function is equal to unity outside this interval i.e. the spacetime becomes
flat. Here also our assumption is that this black hole is formed as a result of
collapse of the spherical null shell which has mass $M$ \cite{28}. This black
hole exists for some time $\Delta V$, and after that it completely
disappears due to the collapse of some other shell which has mass $-M.$ It
is possible to find the gain function for such type of black hole whose
interior spacetime is static. Let us assume that an incoming
radial photon whose initial energy is $E_{1}$, reaches the first shell at
distance $r_{-}$. It moves through the interior spacetime between the
shells and after crossing the second shell at distance $r_{+}$, leaves the
black hole with energy $E_{2}$. We call such a photon of radial type I, and $r_{-},~r_{+}~$ are the points of entrance and exit of photon. Then the gain
function is given by 
\begin{equation}
\chi =\frac{f_{-}}{f_{+}}=\frac{dr_{-}}{dr_{+}},  \label{d3}
\end{equation}
where $f_{\pm }$ are the quantity $f$\ evaluated at $r_{\pm }$. For the
photons which propagate along the horizon $r_{+}=r_{-},$ one gets $f_{-}=f_{+}=0.$ Therefore 
\begin{equation}
\chi _{H}=\exp \left( -\kappa _{H}p\right) .  \label{d4}
\end{equation}
Now, consider type II, a beam of incoming radial photons having energy $E_{1}$. The radial type II photons cross the first shell in the interval $\left( r_{-},r_{-}+\Delta r_{-}\right) ,$ and then the second shell between the
interval $\left( r_{+},r_{+}+\Delta r_{+}\right) ~$having energy $E_{2}$. Then 
\begin{equation}
E_{-}\Delta r_{-}=E_{-}\frac{dr_{-}}{dr_{+}}\Delta r_{+}=E_{-}\frac{f_{-}}{
f_{+}}\Delta r_{+}=E_{+}\Delta r_{+}.  \label{d5}
\end{equation} 
The rate of energy flux is then given by 
\begin{equation}
\xi _{\pm }=\frac{E_{\pm }}{\Delta r_{\pm }},  \label{d6}
\end{equation} 
so that we get 
\begin{equation}
\xi _{+}=\chi ^{2}\xi _{-}.  \label{d7}
\end{equation} 
Since the radial type II photon starts motion in the interval $(0,p),$ the
gain function can be obtained as 
\begin{equation}
\varkappa =\exp \left[ \frac{pq^{\frac{1}{2}}}{2^{\frac{3}{4}}\beta ^{\frac{1}{4}}}\left( \frac{8x^{3}\beta ^{\frac{3}{2}}\exp \left( -\frac{1}{x^{4}}
\right) +2x\beta ^{\frac{1}{2}}l^{2}\Gamma \left( \frac{1}{4},\frac{1}{x^{4}}
\right) ^{2}-2x^{4}\beta ^{\frac{3}{2}}\Gamma \left( \frac{1}{4},\frac{1}{x^{4}}\right) }{2^{\frac{5}{2}}x^{6}\beta ^{2}+\sqrt{2}l^{4}\Gamma \left( 
\frac{1}{4},\frac{1}{x^{4}}\right) ^{2}+2^{\frac{5}{2}}l^{2}\beta
x^{3}\Gamma \left( \frac{1}{4},\frac{1}{x^{4}}\right) }\right) \right] ,
\label{d8}
\end{equation} 
where we use (\ref{A22}). For the double shell model, having a constant metric in the interior, the gain function is given by 
\begin{equation}
\chi =\frac{1}{f_{+}},  \label{d9}
\end{equation} 
and $f$\ $\left( r\right) $ is given by Eq. (\ref{d2}). For type III null
rays, that is, those rays which are outside the interval $(0,p)$ the gain
function is equal to 1.

The quantum radiation from magnetized non-singular black hole can be
estimated with the help of a result from Ref.  \cite{29}, where conformal anomaly is used to work out two-dimensional quantum average of the energy-momentum tensor. Now, massless particles are created from the initial
vacuum state. So for type I rays, the rate of energy flux of these massless
particles is given by the following expression 
\begin{equation}
\xi =\frac{1}{192\pi }\left[ -2\frac{d^{2}\digamma }{dr_{+}^{2}}+\left( 
\frac{d\digamma }{dr_{+}}\right) ^{2}\right] ,  \label{d10}
\end{equation}%
where $\digamma =\ln \left\vert \chi \right\vert .$ Using (\ref{d3}) this
takes the form 
\begin{equation}
\xi =\frac{1}{192\pi }\frac{G\left( r_{-}\right) -G\left( r_{+}\right) }{%
f^{2}\left( r_{+}\right) },~~r_{-}=r_{-}\left( r_{+}\right) ,  \label{d11}
\end{equation}%
where the function $G$ is introduced as 
\begin{equation}
G\left( r\right) =-2f\left( r\right) \frac{d^{2}f}{dr^{2}}+\left( \frac{df}{%
dr}\right) ^{2}.  \label{d12}
\end{equation}%
Again using Eq. (\ref{A22}) we get the result%
\begin{equation}
\xi =\left( \frac{2}{\beta q^{2}}\right) ^{\frac{1}{2}}\frac{1}{192\pi
f^{2}\left( x_{+}\right) }\left[ -2f\left( x_{-}\right) \frac{d^{2}f}{%
dx_{-}^{2}}+\left( \frac{df}{dx_{-}}\right) ^{2}+2f\left( x_{+}\right) \frac{%
d^{2}f}{dx_{+}^{2}}-\left( \frac{df}{dx_{+}}\right) ^{2}\right] .
\label{d13}
\end{equation}%
Here $f\left( x\right) $ and its first and second order derivatives are
given by Eqs. (\ref{A23}), (\ref{A25}) and (\ref{A26}), respectively. The
values $x_{-}$ and $x_{+}$ are related to $r_{-}$ and $r_{+}$ where the
photon crosses the first and second shells.

For type II rays, i.e., when $0<v<p$, the corresponding outgoing null ray
intersects the second shell only, with negative mass, and one has $f\left(
r_{-}\right) =1,$ so that the expression for the energy flux becomes%
\begin{equation}
\xi =\frac{1}{192\pi }\left( \frac{2}{\beta q^{2}}\right) ^{\frac{1}{2}}%
\frac{2f\left( x_{+}\right) \frac{d^{2}f}{dx_{+}^{2}}-\left( \frac{df}{dx_{+}%
}\right) ^{2}}{f^{2}\left( x_{+}\right) }.  \label{d14}
\end{equation}%
For type III rays we get $\xi =0$, because in that region there are no quantum radiations.

\section{Conclusion}

There are different approaches for studying black holes in the framework of
nonlinear electrodynamics. In this letter we used the model of exponential
electrodynamics which is reducible to Maxwell electrodynamics in the weak
field limit. In this model briefriengence phenomenon holds and also weak
energy condition is satisfied, and similarly, causality and unitarity
principles are also satisfied. A regular black hole solution has been obtained in exponential nonlinear electrodynamics in the framework of Einstein's general relativity \cite{13}. Using this method we coupled the model of exponential
electrodynamics with some UV regulating modified theory of gravity to obtain the
asymptotic magnetically charged non-singular black hole metric. The
resulting metric is regular at the origin as the curvature invariants are
finite there. Moreover, the asymptotic values of the metric functions at $r\rightarrow 0$ and $r\rightarrow \infty $ have been worked out. This metric
is similar to the electrically charged non-singular metric \cite{27}.
Further we calculated its thermodynamical quantities such as Hawking
temperature and heat capacity. We note that the first order phase transition occurs at
the outer horizon $r_{2}=\pm 2^{\frac{-1}{4}}\beta ^{\frac{1}{4}}\sqrt{q}
5\times 10^{29},$ as Hawking temperature is zero at these values, for
particular values of $\beta ,q$ and $l$. Here $r_{2}<-0.050194\sqrt{q}\beta
^{\frac{1}{4}},$ the heat capacity is negative so the black hole is
unstable. The heat capacity diverges as the Hawking temperature increases, so
the second order phase transition occurs in the interval $x_{2}\in \left(
-0.05969,0.05969\right)$. Furthermore, quantum radiations from this
non-singular black hole have also been discussed. Here we find the expression for
the quantum energy flux of the massless particles emitted from the interior
of such a non-singular black hole. 

In the limit $\beta \rightarrow 0$, our results correspond to the case of magnetically charged non-singular black holes where Maxwell's electrodynamics is coupled with the modified gravity \cite{28}. For $l=0$, these results coincide with the case of black hole solutions of Einstein's gravity in the presence of exponential electromagnetic field \cite{13}. When both the parameters $\beta$ and $l$ vanish, our results reduce to those for the Reissner-Nordstr\"{o}m black hole of Einstein's theory. Similarly, neutral black hole solutions are obtained if we put $q=0$ and $l=0$ in the formulae derived in this letter.

\section*{Acknowledgements}
Research grants from the Higher Education Commission of Pakistan under its Project Nos. 20-2087 and 6151 are gratefully acknowledged.


\begin{thebibliography}{99}
\bibitem{1} M. Born and L. Infeld, Proc. R. Soc. London A \textbf{144}
(1934) 425.

\bibitem{2} S. I. Kruglov, J. Phys. A \textbf{43}  (2010) 375402
(arXiv:0909.1032).

\bibitem{3} B. Hoffmann, L. Infeld, Phys. Rev. \textbf{51} (1937) 765.

\bibitem{4} A. Peres, Phys. Rev. \textbf{122 } (1961) 273.

\bibitem{5} R. Pellicer and R. J. Torrence, J. Math. Phys. \textbf{10} (1969) 1718.

\bibitem{6} J. D. Jackson, \textit{Classical Electrodynamics, Second Ed., }
John Wiley and Sons, (1975).

\bibitem{7} N. Breton, Phys. Rev. D \textbf{67} (2003) 124004 
(arXiv:hep-th/0301254).

\bibitem{8} S. H. Hendi, Ann. Phys. \textbf{333} (2013) 282 
(arXiv:1405.5359).

\bibitem{9} L. Balart and E. C. Vagenas, Phys. Rev. D \textbf{90} (2014) 124045
 (arXiv:1408.0306).

\bibitem{10} S. I. Kruglov, Int. J. Geom. Meth. Mod. Phys. \textbf{12}
(2015) 1550073  (arXiv:1504.03941).

\bibitem{11} S. I. Kruglov, Ann. Phys. (Berlin) \textbf{528} (2016) 588 
(arXiv:1607.07726).

\bibitem{12} S. I. Kruglov, Phys. Rev. D \textbf{94} (2016) 044026 
(arXiv:1608.04275).

\bibitem{13} S. I. Kruglov, Ann. Phys. \textbf{378} (2017) 59.

\bibitem{14} K.S. Stelle, Phys. Rev. D \textbf{16} (1977) 953.

\bibitem{15} T. Biswas, E. Gerwick, T. Koivisto and A. Mazumdar, Phys. Rev.
Lett. \textbf{108} (2012) 031101, arXiv:1110.5249 [gr-qc].

\bibitem{16} L. Modesto and L. Rachwal, Nucl. Phys. D \textbf{889} (2014) 228, arXiv:1407.8036 [hep-th].

\bibitem{17} S. Talaganis, T. Biswas and A. Mazumdar, Class. Quant. Grav. 
\textbf{32} (2015) 215017 , arXiv:1412.3467 [hep-th].

\bibitem{18} E. T. Tomboulis, ``Nonlocal and quasi-local field theories,"
(2015), arXiv:1507.00981 [hep-th].

\bibitem{19} E. T. Tomboulis, Mod. Phys. Lett. A \textbf{30} (2015) 1540005.

\bibitem{20} E. Spallucci, A. Smailagic and P. Nicolini, Phys. Rev. D 
\textbf{73} (2006) 084004 , arXiv:hep-th/0604094 [hep-th].

\bibitem{21} T. Biswas, T. Koivisto and A. Mazumdar, JCAP \textbf{1011}
(2010) 008, arXiv:1005.0590 [hep-th].

\bibitem{22} L. Modesto, W. M. John and P. Nicolini, Phys. Lett. B 
\textbf{\ 695} (2011) 397,  arXiv:1010.0680 [gr-qc].

\bibitem{23} S. Hossenfelder, L. Modesto and I. P. Schwarz, Phys. Rev. D 
\textbf{81} (2010) 044036, arXiv:0912.1823 [gr-qc].

\bibitem{24} A. Conroy, A. Mazumdar and A. Teimouri, Phys. Rev. Lett. 
\textbf{114} (2015) 201101, arXiv:1503.05568 [hep-th].

\bibitem{25} M. Markov, JETP Letters \textbf{36} (1982) 265.

\bibitem{26} M. Markov, Ann. Phys. \textbf{155} (1984) 333.

\bibitem{27} V. P. Frolov, Phys. Rev. D \textbf{94} (2016) 104056, arXiv:1609.01758 [gr-qc].

\bibitem{28} V. P. Frolov and A. Zelnikov, Phys. Rev. D \textbf{95} (2017) 044042, arXiv:1612.05319 [hep-th]. 

\bibitem{SH} S. W. Hawking, Commun. Math. Phys. \textbf{43} (1975) 199.

\bibitem{GS} U. A. Gillani and K. Saifullah, Phys. Lett. B \textbf{699} (2011) 15. 

\bibitem{RS} M. Rehman and K. Saifullah, JCAP \textbf{03} (2011) 001. 

\bibitem{29} S. M. Christensen and S. A. Fulling, Phys. Rev. D 
\textbf{15} (1977) 2088. 

\end{thebibliography}
\end{document}